\documentclass[journal,twocolumn,11pt]{IEEEtran}
\usepackage{amsmath,amsfonts}
\usepackage{algorithmic}
\usepackage{array}
\usepackage[caption=false,font=normalsize,labelfont=sf,textfont=sf]{subfig}
\usepackage{textcomp}
\usepackage{stfloats}
\usepackage{url}
\usepackage{color}
\usepackage{verbatim}
\usepackage{graphicx}
\usepackage{tikz}
\usepackage[usestackEOL]{stackengine}

\usetikzlibrary{shapes.arrows,decorations.pathreplacing,decorations.markings,shapes.geometric}
\tikzset{radiation/.style={{decorate,decoration={expanding waves,angle=0,segment length=5pt}}}} 

\def\BibTeX{{\rm B\kern-.05em{\sc i\kern-.025em b}\kern-.08em
    T\kern-.1667em\lower.7ex\hbox{E}\kern-.125emX}}
\usepackage{balance}
\begin{document}
\title{Physical Layer Security - from Theory to Practice}
\author{Miroslav Mitev, Thuy M. Pham, Arsenia Chorti, André Noll Barreto,  Gerhard Fettweis}

\maketitle

\begin{abstract}

A large spectrum of technologies are collectively dubbed as physical layer security (PLS), ranging from wiretap coding, secret key generation (SKG), authentication using physical unclonable functions (PUFs), localization / RF fingerprinting, anomaly detection monitoring the physical layer (PHY) and hardware. Despite the fact that the fundamental limits of PLS have long been characterized, incorporating PLS in future wireless security standards requires further steps in terms of channel engineering and pre-processing. Reflecting upon the growing discussion in our community, in this critical review paper, we ask some important questions with respect to the key hurdles in the practical deployment of PLS in 6G, but also present some research directions and possible solutions, in particular our vision for context-aware 6G security that incorporates PLS. 
\end{abstract}



\section{Introduction}

In 1949, Shannon introduced the concept of perfect secrecy~\cite{ShannonEntropyIntro1949} and demonstrated that xor-ing a message $\mathbf{m}$ with a uniform random key $\mathbf{k}$ of the same length to obtain a ciphertext $\mathbf{c=m \oplus k}$, provides perfect secrecy, i.e., for one-time pad schemes it can be shown that $\mathbb{H}(\mathbf{m}| \mathbf{c})=\mathbb{H}(\mathbf{m})$, where $\mathbb{H}(\cdot)$, denotes entropy. Although the one-time pad is impractical, it showcases that randomness to induce equivocation is a cornerstone of confidentiality, i.e., given enough confusion at the adversarial end, provably unbreakable crypto systems can be developed. 

This idea forms the basis of PLS and in particular of the wiretap coding. In Wyner's pioneering work~\cite{Wyner1975}, it was demonstrated that excess noise in the link to an eavesdropper can be exploited for keyless transmission of confidential messages, while guaranteeing reliability. For additive white Gaussian noise (AWGN) channels~\cite{Leung:Gaussian:1978}
and the general class of symmetric channels~\cite{bloch_barros_2011},
the maximum rate at which both reliability and confidentiality can be simultaneously guaranteed, referred to as the secrecy capacity, is equal to the excess capacity of the legitimate link with respect to the eavesdropper's link. A few years later this idea was generalized to the broadcast wiretap channel by Csisz\'ar and K\"orner~\cite{Csiszar1978}.

Since then, the idea of exploiting entropy sources at PHY to achieve specific security goals has been extensively researched~\cite{bloch_barros_2011,Poor2009,Maurer};
apart from confidentiality using wiretap coding, opportunities for key generation and distribution, user and device authentication and resilience to PHY denial of service attacks have been identified.

\textcolor{black}{A mature research direction is that of secret key generation (SKG) from a common random source. Given the observations of this source both by authorized users and by an eavesdropper, the fundamental limits on the key generation rates were derived in~\cite{Maurer}. In communications, especially wireless, the propagation channel itself can be such a random source, allowing it to be used to distill secret keys, which can be used for pairing and encryption.} The corresponding procedures are well studied and
 numerous practical demonstrators have been developed~\cite{Zhang_Access2020}
 along with
and concrete countermeasures in the case of active attacks~\cite{Belmega_TIFS2017, Mitev_Entropy2021}. 

With respect to authentication, key approaches include physical unclonable functions (PUFs), localization-based authentication and RF fingerprinting. PUFs exploit the unclonable variability in hardware manufacturing processes for authentication, while localization and RF fingerprinting are widely used soft authentication factors~\cite{Mitev_Access_2022}. 

Integrating the above mentioned technologies into communication systems comes with the promise of a new breed of lightweight, quantum resilient, low-latency, low-footprint and scalable security schemes. However, after decades of research, the deployment of practical PLS solutions is still in its infancy and has met significant resistance. In this paper, we first discuss whether a fundamental change in security is actually necessary in order to ensure trustworthy future generations and will try to show the importance of PHY when evaluating trust in future wireless networks. Then, we will discuss some of the key reasons behind the lag between theory and practice in PLS, and propose a roadmap to bridge the gap between theoretical analysis to products. Finally, we will present our more general vision for an intelligent, context-aware 6G security, incorporating the physical layer for the first time.

The rest of the paper is organized as follows. In Section II we discuss the reasons why PLS is pertinent to 6G. Section III presents the current state-of-the-art in PLS, along with open research issues, while Section IV presents future perspectives and concludes the paper. 

\section{Trustworthy and Resilient 6G and the Role of the Physical Layer}

The sixth generation of wireless will interconnect intelligent and autonomous cyberphysical systems, like robots, drones, vehicles, platoons, etc. In this emerging “fusion” of the digital and physical worlds, standard authentication and access control schemes do not suffice to build trust and evaluate the trustworthiness of autonomous agents. In essence, building a trustworthy and resilient 6G boils down to trusting:
\begin{enumerate}
\item The autonomous multi-agents; 
\item Their sensing inputs (that drive their decisions); 
\item The communication links between them; 
\item The computations performed (including learning and optimization);
\end{enumerate}

Until recently, trust for the autonomous agents has primarily focused on the trustworthiness and explainability of the artificial intelligence algorithms that govern them, e.g., using coalitional game theory tools such as Shapley values, evidence theory, etc.\cite{Liang_Nature2022}.
At the same time, reputation-based and crowd-vetting approaches have been widely investigated, e.g., \cite{Gong-Poor2015,Mallmann-Trenn2022}.

A game changer in this area is that it has been recently shown that anomalies in the behaviour of cyberphysical agents can be actually inadvertently identified from behavioural aspects; first to be explored is naturally related to agent positioning. As an example in \cite{Gil-Stephanie_autonRobots2017}, the angle of arrival has been used to identify Sybil attacks in robotic systems, while in the same direction range estimation has been used in \cite{Mitev_Access_2022} to provide resilience against more general impersonation attacks. This direction of research, hinges to the potential incorporation of PLS-based authentication approaches in trust measures for autonomous agents in 6G. Opportunities to provide not only high data rates, but also high-precision ranging and localization to enhance trust need to be systematized by our community.

With respect to trustworthy computation, a key aspect has to do with decentralization, e.g., blockchain technologies, federated learning, crowd-sourcing, private computation~\cite{Trust_federatedLearning-TNNLS2022,Trust_crowdsourcing-WC2016,Trust_informationRetrieval-TIT2020} and private information retrieval are among the technologies currently explored~\cite{Trust_blockchain-IoT2022}, in conjunction with isolation and composability of hardware platforms. Up to now, evaluating the trustworthiness of computation is a task perceived to belong entirely to the digital domain. It remains to be seen whether hardware monitoring will in the future allow to identify untrustworthy computation and importantly help recognise the existence of backdoors in hardware originating from untrusted vendors. 

Challenges also arise to securing the sensing layer itself and rendering it resilient to denial of service and man-in-the-middle attacks. Aspects related to distributed anomaly detection in software defined wireless sensor networks~\cite{Nunez-Chorti_IOT2022} have demonstrated that it is possible in large scale IoT networks to monitor hardware behaviour (memory usage, power consumption, Tx/Rx times, etc.) to identify compromised or faulty sensors. Exploring further aspects including passive and active attacks to sensing, along with related privacy concerns is paramount for a trustworthy 6G.  

Finally, the links between autonomous cyberphysical agents will be vital to determine their behaviour, e.g., in the case of platooning. To this end, unarguably, the security protocols of fifth generation systems are a significant improvement with respect to LTE, resolving many, albeit not all, open issues in older generations of wireless. In particular, securing wireless links under overly aggressive latency constraints, scaling authentication and key distribution to massive numbers to accommodate massive Internet of things (IoT) while providing quantum resistance for constrained devices, persist as open challenges at present, despite recent standardization of four post-quantum cryptographic algorithms from NIST. To address all of these issues, PLS technologies emerge as competitive alternatives or complementary schemes to standard cryptography. 

\begin{figure*}[!t]
    \centering
    \includegraphics[width=0.8\textwidth]{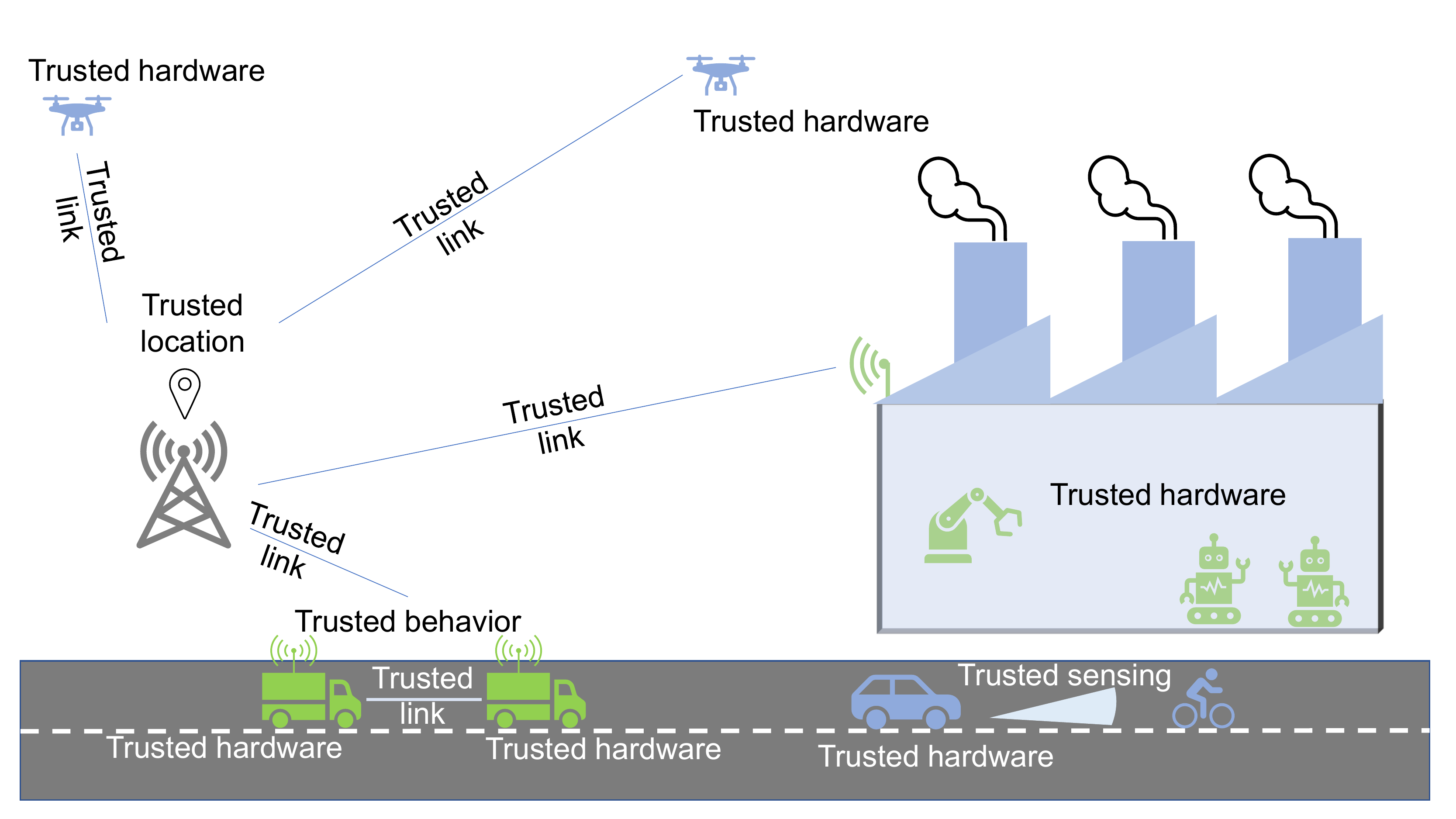}
    \caption{Trustworthy 6G - the role of PLS.}
    \label{fig:trust_6G}
\end{figure*}

We have showcased that 6G trustworthiness needs to include trust of the physical world and infrastructure across the board. A glimpse towards some of the security features that PLS can bring into the 6G world is given in Fig. \ref{fig:trust_6G}. The figure illustrates that physical aspects, e.g., hardware, location, link, behavior, sensing, could bring an additional (and important) asset of properties that could help in ensuring trustworthiness in 6G.
In the following sections, we focus entirely on the trsutworthiness of the communications links. In particular, delving deeper in PLS, we provide an overview of the state-of-the-art and explain how current limitations can be overcome to fulfill the need, as well as the promise, for security controls at all layers, including at the physical layer, for the first time in 6G.

\section{PLS - State-of-the-art and Open Issues}

In this section, we will review only some of the key contributions in the PLS literature.
More importantly, we will identify key open issues that should be addressed before practical deployment.

\subsection{Keyless transmission of confidential messages}
The interest in PLS research is motivated by two pioneering works by Shannon and Wyner who introduced the concepts of perfect secrecy and wiretap channel, respectively \cite{ShannonEntropyIntro1949,Wyner1975}. In \cite{ShannonEntropyIntro1949}, Shannon considered a noiseless system in which a transmitter -- referred to as Alice -- sends a coded message to a receiver -- referred to as Bob -- under the constraint of keeping it confidential from an eavesdropper -- referred to as Eve. He has proven that this is possible by a transformation that generates codewords in the null space of Eve's observations, a condition regarded to as perfect secrecy and which can be fulfilled by a one-time-pad scheme as long as the key entropy is larger than the message entropy. 

In reality however, wireless links are noisy. Thus, Wyner extended the scheme to a more realistic system model by considering a discrete memoryless channel \cite{Wyner1975}. Based on this model, he derived the secrecy capacity for the case of degraded wiretap channels, which was later generalized to the non-degraded case by Csisz\'ar and K\"orner \cite{Csiszar1978}.
The secrecy capacity region, under the assumption of perfect CSI knowledge at the transmitter, has been  characterized for different setups, including multiple input multiple output (MIMO) scenarios~\cite{SDoF_delayedCSI-TIT2013}. 


Furthermore, the concept of secrecy degrees of freedom (SDoF) has been introduced as an alternative metric to simplify calculations~\cite{Poor_EURASIP2009}. Using the SDoF, another important conclusion was made: achieving perfect secrecy when only imperfect CSI is available is possible only when asymmetric statistical properties are present for the channels towards both receivers. In this sense, when the channels have symmetrical properties, positive SDoF can be ensured by paying the cost of additional overhead in terms of side information used to introduce asymmetry at the encoder~\cite{SDoF_delayedCSI-TIT2013}. Having this result, it is clear that the quality of the CSI can play a vital role on the achievable secrecy.

In this regard, an important result has been published in~\cite{MaddahAli2010OnTD} showing that even an outdated CSI at the transmitter can be used towards increasing the SDoF. The general idea is that, delayed CSI can be successfully incorporated towards interference alignment between users. While these are encouraging findings, further research is still needed to render such secrecy mechanisms possible in a more general context. We note in passing that the idea of artificial noise injection has attracted a lot of attention. However, it seems unlinkely that such approaches will be used in practice, at least in the near future, due to strict regulations for the levels of electromagnetic radiations and the need for lowering energy consumption across the board.
Another critical aspect is the availability of the eavesdropper's CSI at the transmitter, which is highly unlikely in many actual scenarios.
To overcome such difficulties, one possible metric is the secrecy outage probability (SOP), which is given by
\begin{equation}
	\mathcal{P}_{out}(R) = \mathcal{P}(C_{S}<R),
\end{equation} where $C_{S}$ denotes the secrecy capacity and $R$ denotes a target secrecy rate. 
Closely related, is the probability of nonzero secrecy capacity, defined as 
\begin{equation}
\mathcal{P}_{NZ} = \mathcal{P}(C_{S}>0) = 1- \mathcal{P}_{SOP}(R=0).
\end{equation}

\begin{figure}
    \centering
    \includegraphics[width=0.5\textwidth]{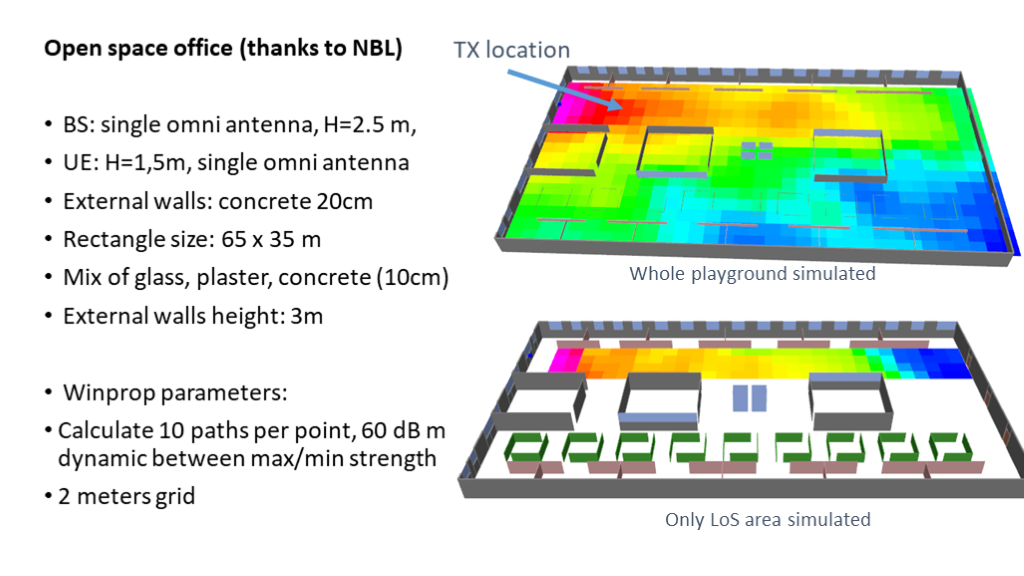}
    \caption{Raytracing based received signal strength evaluation in an indoor environment (courtesy of Nokia Bell Labs Nozay).}
    \label{fig:ray tracing}
\end{figure}

In \cite{Shulz21} it was shown that even in THz systems, weak directivity results in large insecure areas, and while in \cite{Absolute_security-Medard2022} it is shown that although
such areas can be minimized they cannot be fully eliminated. Therefore, at the moment even with ultra-massive MIMO systems and pencil sharp beamforming, it remains an open question how to guarantee zero information leakage without any assumptions regarding the adversarial position, the numbers of antennas, cooperation between distributed adversarial actors, etc. Partially controllable channels, e.g., using intelligent reflective surfaces, could be worth investigated in this aspect to facilitate channel engineering.

Furthermore, efficient CSI estimation is key for wiretap coding; an example using ray-tracing tools is depicted in Fig. \ref{fig:ray tracing}. This could be propelled in 6G by online learning together with location-based channel estimation. Extensive related works have already appeared for mmWave and THz bands.  

\begin{figure}
    \centering
    \includegraphics[width=0.5 \textwidth]{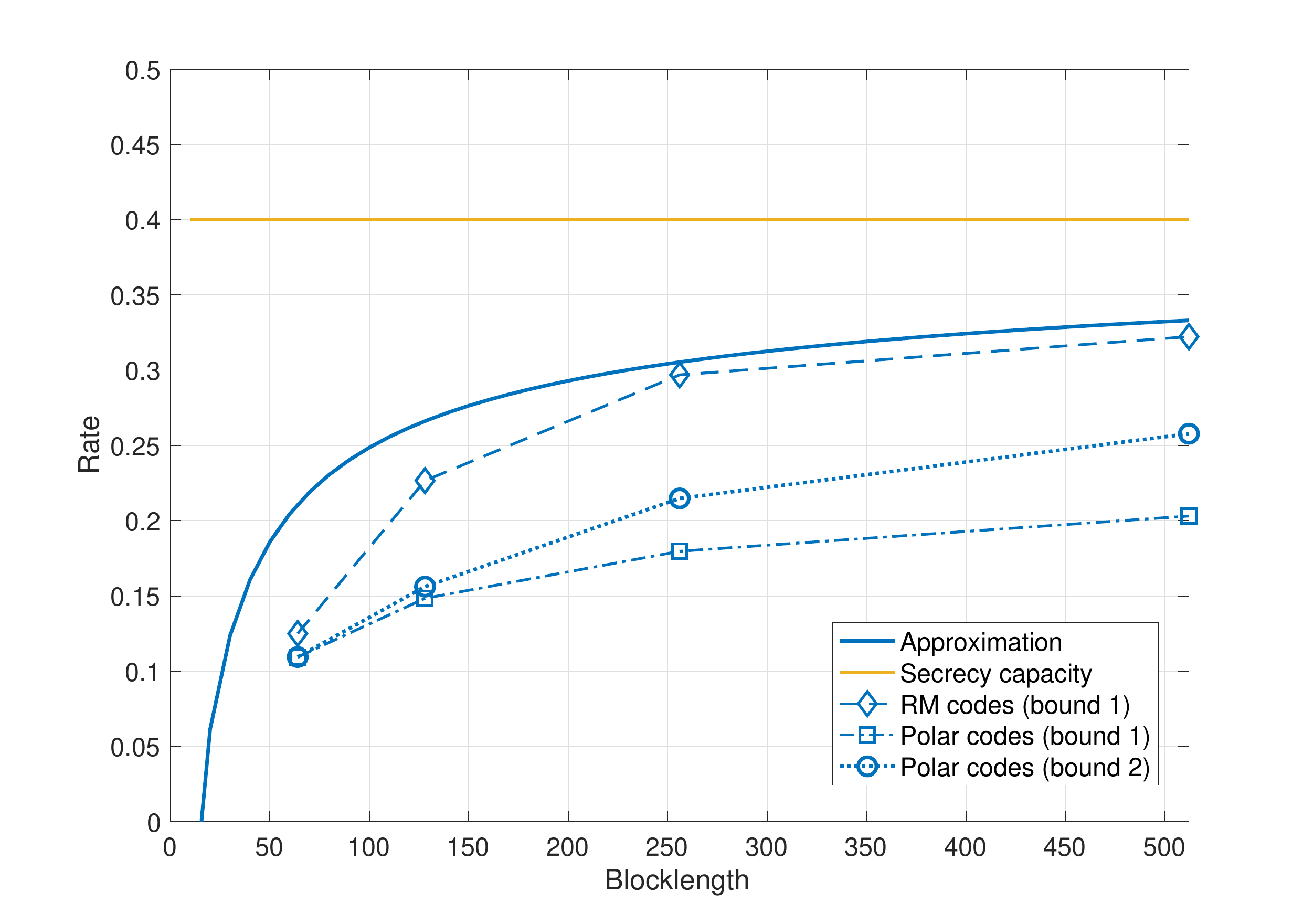}
  \caption{Comparison of the lower bound on achievable secrecy rates of Reed-Muller and polar codes for $p=0.4$ and $\delta=0.001$, with the second order approximation secrecy rate.}
    \label{fig:BEC}
\end{figure}

A further issue concerns the security guarantees in the finite blocklength~\cite{Yang:secrecy:reliability:2017},
as opposed to asymptotic results or very special channel models~\cite{Andersson:NestedPolar:2010,Mahdavifar:PolarCode:2011}.
In \cite{Yang:secrecy:reliability:2017} the achievable secrecy rate was shown to be a function of (i) the blocklength, (ii) the error rate and (iii) the information leakage, i.e., at finite blocklengths it is impossible to guarantee zero information leakage. In Fig. \ref{fig:BEC}, a comparison is provided between the lower bound on the achievable secrecy rates of Reed-Muller and polar codes for a semi-deterministic wiretap channel where the main
channel is noiseless and the wiretap channel is a binary-erasure channel with
erasure probability $p=0.4$ and information leakage $\delta=0.001$, with the second-order approximation secrecy rate~\cite{Shakiba-ITW2021}.

Despite the above negative remarks, wiretap coding could be key among PLS security schemes for 6G. Although strong confidentiality guarantees seem to be unattainable in realistic propagation scenarios due to either radiation leakage in mMIMO arrays, imperfect CSI estimation, or information leakage in finite blocklengths, we can envision its use for privacy purposes, geofencing or for statistical based measures of trust in wireless links. 

Indeed, adaptive security controls are needed for future, highly heterogeneous systems. 
By adaptive security we describe a dynamic security engine, always aiming at
delivering the best security possible in a given context. A framework for the development of related schemes is offered by quality of security (QoSec), which describes the “degree of security” in measurable manner \cite{Context_aware_6g_security}. We argue that a more holistic view on QoSec is needed, incorporating wiretap coding for privacy.


\subsection{Secret key generation (SKG)}

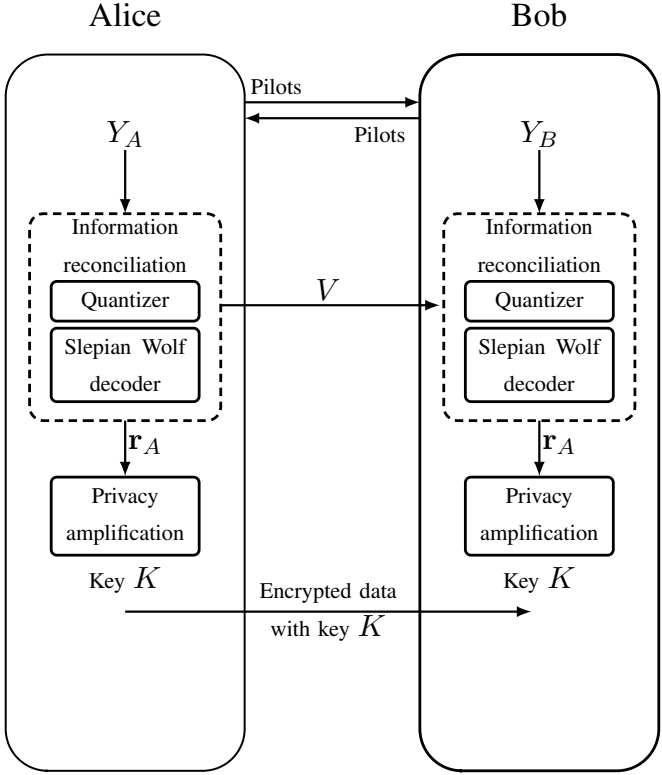
\begin{figure}[t]
\centering
\tikzset{>=latex}
 \resizebox{9cm}{!}{
\begin{tikzpicture}
\begin{scope}[xshift=2cm]
\draw[rounded corners=15pt,line width=0.71pt] (0.5,0)--(3,0)--(3,9)--(0,9)--(0,0)--(.5,0);
\node at (1.5,9.5) {\large Alice};
\draw[thick, ->] (3,8.4) -- (5.2,8.4);
\draw[thick, ->] (5.2,8.2)--(3,8.2);
\node at (3.4,8.6) {\scriptsize Pilots};
\node at (4.7,8) {\scriptsize Pilots};
\node at (1.5,8) {${Y}_A$};
\draw[thick, ->] (1.5,7.8)--(1.5,7);
\node[draw, rounded corners=2pt,line width=1pt, text width=1.6cm, align=center] at (1.5,5.9) {\Longstack[c]{\scriptsize Quantizer}};
\node[draw, rounded corners=2pt,line width=1pt, text width=1.6cm, align=center] at (1.5,5.1) {\Longstack[c]{\scriptsize Slepian Wolf \\ \scriptsize decoder }};
\draw[rounded corners=5pt,line width=1pt,densely dashed] (1,4.4)--(2.7,4.4)--(2.7,7)--(0.3,7)--(0.3,4.4)--(1,4.4);
\node at (1.5,6.6) {\Longstack[c]{\scriptsize Information \\ \scriptsize reconciliation}};
\draw[thick, ->] (2.7,5.85)--(5.45,5.85);
\node at (4.05,6.05) {$V$};
\draw[thick, ->] (1.5,4.4)--(1.5,3.7);
\node at (1.75,4.1) {$\mathbf{r}_A$};
\node[draw, rounded corners=2pt,line width=1pt, text width=1.6cm, align=center] at (1.5,3.2) {{\scriptsize Privacy \\ \scriptsize amplification }};
\node at (1.5,2.4) {\scriptsize Key \normalsize  ${K}$};
\draw[thick, ->] (1.5,2)--(6.6,2);
\node at (4.05,2) {\Longstack[c]{\scriptsize Encrypted data  \\ \scriptsize with key \normalsize $K$}};
\end{scope}
\begin{scope}[xshift=7.2cm]
\draw[rounded corners=15pt,line width=1pt] (0.5,0)--(3,0)--(3,9)--(0,9)--(0,0)--(.5,0);
\node at (1.5,9.5) {\large Bob};
\node at (1.5,8) {${Y}_B$};
\draw[thick, ->] (1.5,7.8)--(1.5,7);
\node[draw, rounded corners=2pt,line width=1pt, text width=1.6cm, align=center] at (1.5,5.9) {\Longstack[c]{\scriptsize Quantizer}};
\node[draw, rounded corners=2pt,line width=1pt, text width=1.6cm, align=center] at (1.5,5.1) {\Longstack[c]{\scriptsize Slepian Wolf \\ \scriptsize decoder }};
\draw[rounded corners=5pt,line width=1pt,densely dashed] (1,4.4)--(2.7,4.4)--(2.7,7)--(0.3,7)--(0.3,4.4)--(1,4.4);
\node at (1.5,6.6) {\Longstack[c]{\scriptsize Information \\ \scriptsize reconciliation}};
\draw[thick, ->] (1.5,4.4)--(1.5,3.7);
\node at (1.75,4.1) {$\mathbf{r}_A$};
\node[draw, rounded corners=2pt,line width=1pt, text width=1.6cm, align=center] at (1.5,3.2) {\Longstack[c]{\scriptsize Privacy \\ \scriptsize amplification }};
\node at (1.5,2.4) {\scriptsize Key \normalsize  $K$};
\end{scope}
\end{tikzpicture}
}
\caption{Secret key generation between Alice and Bob. Thanks to reciprocity, the quantizer outputs can be expressed as
\begin{math}
\mathbf{r}_{A}=\mathbf{d}\oplus\mathbf{e}_{A},
\mathbf{r}_{B}=\mathbf{d}\oplus\mathbf{e}_{B}.
\end{math} Using $V$ Bob corrects the errors to obtain $\mathbf{r}_{A}$.} \label{fig:Miro SKG}
\end{figure}

Typically, the SKG process consists of three phases~\cite{Maurer,Bennett95},
\cite{Mitev_EURASIP2020} as depicted in Fig. \ref{fig:Miro SKG}. 
In the first phase, referred to as \textit{shared randomness distillation}, Alice and Bob observe a common random source, and their observations, denoted by $Y_A, Y_B$, respectively, are dependent random variables. An eavesdropper, referred to as Eve, observes $Y_E$, which may be correlated or not to $Y_A$ and $Y_B$. In wireless channels, a readily available source of shared randomness is the multipath fading, which is caused by reflections, diffraction and scattering from a random environment. In case the same frequency is used, then the equivalent baseband channel between two nodes is reciprocal during the coherence time~\cite{Mukherjee14}.
These observations are typically done by sampling the channel either in time, frequency or both, followed by a quantization process. Although the channel is reciprocal, transceivers at Alice and Bob will have different RF-chain impairments, different noise and interference realisations, and, in TDD systems, they will sample the channel at different instants. This means that their observations will not be exactly the same, and thus \textit{information reconciliation} is performed with the exchange of side information $V$. This step has to necessarily  be followed by \textit{privacy amplification}, in which the key size is adjusted to take into account possible information leakage to the eavesdropper by estimating the conditional min entropy. 

At the end of the SKG process, a common key $K \in \mathcal{K}$ is extracted at Alice and Bob, such that, for any $\epsilon >0$, the following statements hold \cite{Csiszar}:
\begin{eqnarray}
 \text{Pr}\!\left(K\!=\!f_A\left(Y_A, V\right)\!=\!f_B\left(Y_B, V\right)\right)\!\!&\!\!\geq\!\!&\!\! 1-\epsilon ,\label{eq:Ka=Kb}\\
 I(K;V)\!\!&\leq&\!\! \epsilon ,\label{eq:negligible leakage}\\
 H(K)\!\!&\geq&\!\! \log |\mathcal{K}| -\epsilon \label{eq:key uniformity},
 \end{eqnarray}
{where $H(K)$ denotes the entropy of the key $K$ and $I(K;V)$ denotes the mutual information between $K$ and $V$.}

The first inequality demonstrates that the SKG process can be made error free; (\ref{eq:negligible leakage}) ensures that the exchange of side information through public discussion does not leak any information to eavesdroppers; while (\ref{eq:key uniformity}) establishes that the generated keys attain maximum entropy (i.e., are uniform). {Under the three conditions, an upper bound on the rate for the generation of secret keys is given by \cite{Maurer}
\begin{eqnarray}
\min\left\{I(Y_A; Y_B), I(Y_A; Y_B | Y_E) \right\}.
\end{eqnarray}
Assuming rich multipath environments, the decorrelation properties of the wireless channel over short distances can be exploited to ensure that Eve's observation $Y_E$ is uncorrelated with $Y_A$ and $Y_B$~\cite{Ye06,Reznik2010,Chou12,Mukherjee14,Mitev_EURASIP2020};
in this case, the SKG capacity is given by \cite[Sec. II]{Csiszar}}
\begin{math}
C_K=I(Y_A;Y_B). \label{eq:key rate}
\end{math}

However, this condition is rarely met in real life. In particular, correlations and dependencies in four domains, \textit{space, time, frequency and antenna} between Alice's, Bob's and the Eve's observations have to be taken explicitly into account. While subsampling in the time, frequency and antenna domains can constitute simple approaches to re-create a memoryless channel, so that the observations between Alice and Bob are independent from the observations of Eve along these domains, correlations and dependencies in space need on the other hand to be taken explicitly into account. Pre-processing steps to address these issues have recently been reported in \cite{ WSA21,Srinivisan21}. 

Furthermore, active attacks have been addressed in \cite{Belmega_TIFS2017, Mitev_Entropy2021} and hybrid designs of authenticated encryption leveraging SKG along with symmetric block ciphers have appeared in \cite{Mitev_EURASIP2020}. As a result, SKG emerges as one of the most mature and promising PLS technologies for 6G. Clearly, SKG will be helpful in use cases where key distribution is a major issue, such as massive IoT, addressing scalability in constrained devices that cannot run public key encryption handshakes (or their corresponding post-quantum counterparts). 

SKG is a mature technique, but one of its major challenges is that the achievable key generation rate depends on the channel statistics. However, upper layers will require a minimum or at least a known rate. Understanding how the achievable rate depends on the channel parameters was the subject of several papers \cite{MITEV_vtc2022, MITEV_globecom2022},
but is still an open issue. Furthermore, parts of the SKG algorithm itself, like the sampling rate in time and frequency, and the CSI quantizer should also be optimized according to the channel properties.

Another practical issue is that SKG depends on the availability of reliable CSI information. However, existing wireless chipsets usually do not provide this information, and, even, if they did, they would have to be trusted to provide the correct information. An alternative, using a separate encryption box was proposed in \cite{Zoli_EURASIP2020}.

As a final note, other entropy sources can be exploited for key distillation, e.g., by leveraging the sensing layer and can be further exploited for device pairing. 

\subsection{Physical unclonable functions and biometrics}
Some of the most prominent authentication techniques that come from the physical layer are physical unclonable functions (PUFs) and biometrics. The idea of PUFs is to authenticate devices using the unique properties of integrated circuits (ICs). Such properties appear due to unpredictable variations during their fabrication process. To build a protocol, such variations are typically used in a challenge-response manner. Depending on the PUF architecture, a challenge could refer to measuring gate delays, power-on state or other variable features. 

\begin{figure*}[!t]
    \centering
    \includegraphics[clip, trim=2cm 7cm 2cm 1cm,width=0.8\textwidth]{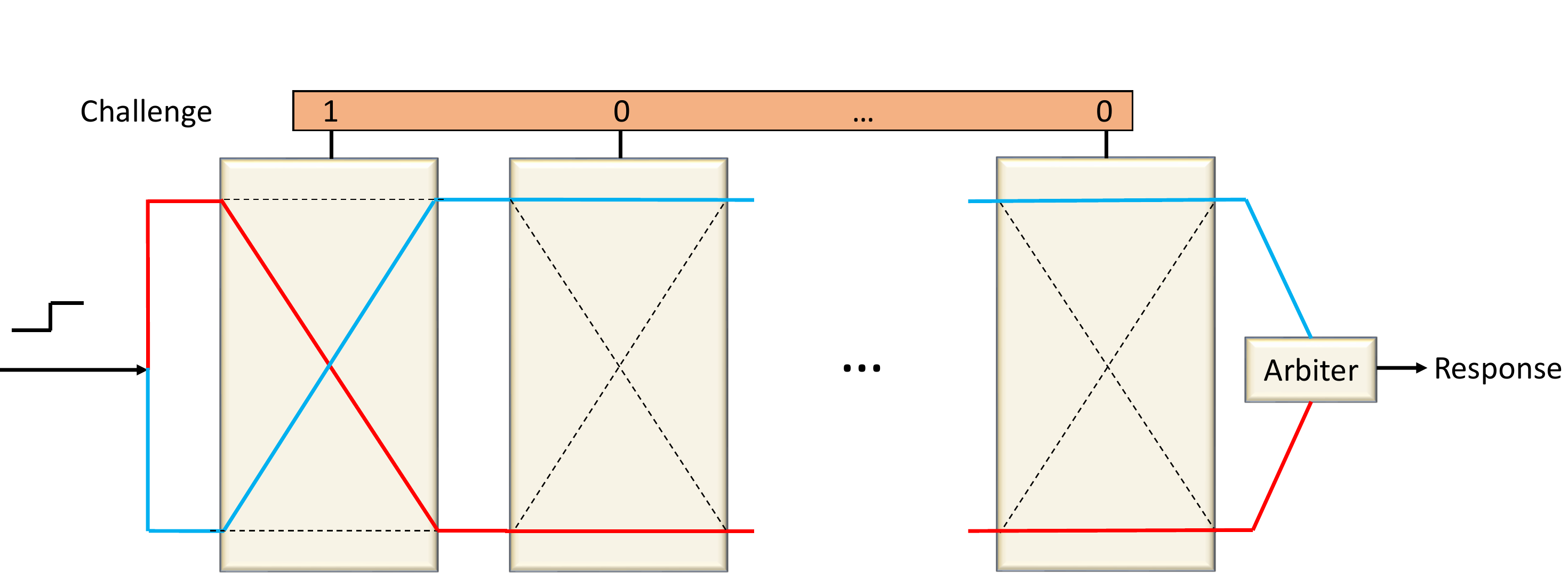}
    \caption{Arbiter PUF.}
    \label{fig:arbiter_puf}
\end{figure*}

A popular architecture, illustrated in Fig. \ref{fig:arbiter_puf}, is the arbiter PUF. The scheme is based on the transmission of rising edge signals through two ``identical'' delay paths, each composed of series of switching elements. Due to variation properties the delay required for each signal to pass through the trace will be different. A challenge to this scheme, as illustrated in  Fig. \ref{fig:arbiter_puf}, is a bit sequence that defines the configuration of the switching elements; and a response is a single bit output that defines which signal arrives first at the end. Depending on the number of challenge-response pairs (CRPs) that a PUF can support, architectures are divided into two groups: weak and strong PUFs. The number of CRPs of a weak PUF increases linearly or polynomially with the component blocks (some architectures support only a single CRP) and the number of CRPs of a strong PUF increases exponentially with the component blocks. In this sense, arbiter PUF is considered to be a strong PUF.

Following from the discussion above, biometrics can be seen as a weak PUF structure that measures  unique birthmarks of human users (as opposed to devices). Such features include voice, palm vein, iris, behavioral biometrics and more. Each of these features can produce a single CRP for user authentication. In this sense, building a PUF-based or a biometric-based authentication protocol requires identical steps:

\textit{Enrollment} -- this step is carried out offline on a secure channel. During enrollment, a set of responses $R_1, \dots, R_i \in \mathcal{R}$ (biometric or PUF) are collected by running a set of challenges $C_1, \dots, C_i \in \mathcal{C}$. Additionally, the measurement noise of the process is characterized in order to generate helper data $hd$. An authenticator creates a database where CRPs and helper data are associated to a particular user/device.

\textit{Authentication} -- during the online authentication step, the authenticator sends a random challenge $C_i$ from its database to the corresponding user requesting to reproduce the response $R_i$. The user then replies with its PUF or biometric measurement $R'_i$. Due to presence of noise the newly generated response will differ from the one generated during enrollment, i.e., $R'_i \neq R_i$, therefore the helper data is used in a reconciliation decoder to regenerate $R_i$, in which case authentication is successful. To prevent replay attacks a CRP pair should never be reused, or other measures should be taken, e.g., time stamps.
Next, some key issues in the application of such authentication approaches are discussed.

First, a topic that is seeing growing interest is the privacy of biometric data.   
To perform biometrics-based authentication, the collected measurements are normally passed through third-party authentication servers. This may lead to privacy leakage, i.e., users are clueless about how and where their data is stored or used. Furthermore, as biometrics are permanent features, if adversaries get access to the collected data they could use it to build a human-digital twin. Therefore, it is important that biometric protection techniques are employed. 
One approach that can be used to avoid storing biometric data is through the use of homomorphic encryption~\cite{homomorphic_biometrics}. In such a scheme, performing an operation on the encrypted data is equivalent to performing the same operation on the plaintext. Hence, users can provide only encrypted biometric data to authenticate themselves without revealing sensitive content. 
However, homomorphic encryption requires complex and slow operations, i.e., it is not suitable for constrained devices and low-latency scenarios. In this sense, further research on lightweight and secure biometric protection is required. 

Another important topic that has to be addressed concerns the unclonability and randomness of PUFs.
First, due to the low number of CRPs supported by weak PUFs, they are susceptible to exhaustive search attacks. 
Strong PUFs, on the other hand, have large CRP space which makes exhaustive search attacks impractical. However, the interactive fashion of executing the authentication protocol described above can leak numerous CPRs and in specific cases helper data streams. 
It has been shown that an attacker can use the leaked information in machine-learning (ML) algorithms to successfully model a PUF~\cite{PUF_study_2021}. Some of the directions that can help solving this issue are the introduction of more complex structures, e.g., 
XOR-ing the outputs of multiple PUFs, as opposed to using their individual outputs, can already prevent multiple ML modelling attacks~\cite{XORing_PUFs}. This gives a shorter but unpredictable sequence. 

Another approach is the combination of different PLS schemes, e.g., PUFs and SKG~\cite{Mitev_Access_2022}. During the authentication procedure both parties can generate a shared secret key. The key can then be used to encrypt and hide the transmission of CRPs and/or helper data, avoiding leakage to adversaries. 
Another issue concerning the uniqueness of PUFs is their initial min-entropy, i.e., the randomness of their outputs. A low min-entropy of a PUF (i.e., there is tendency to produce more 0s or 1s) opens up the chance for statistical inference attacks. Therefore the design of high-entropy PUF architectures is another important research topic. 

Finally, it is important to identify appropriate use cases for both, biometrics and PUFs. There are already a variety of commercial products for both PUF~\cite{PUF_commercial1} and biometric authentication~\cite{Biometric_commercial}, however,  
as noted above, when either of the techniques is used as a single authentication factor there might be serious concerns. Therefore, the combination of PUF, biometrics and other authentication factors can be used towards building a secure and reliable multi-factor authentication. Scenarios where such approach might be beneficial include eHealth (e.g., for accessing medical records), smart factories (e.g., for access control) and commercial applications (e.g., online banking). 
What is important to mention is that these schemes are not here to replace authentication handshakes but to contribute for their efficient and lightweight implementation.



\subsection{Location-based authentication and RF fingerprinting}
Apart from PUF authentication, there are other PHY-based authentication techniques which can be categorized into two types, i.e., RF-based and location-based. 

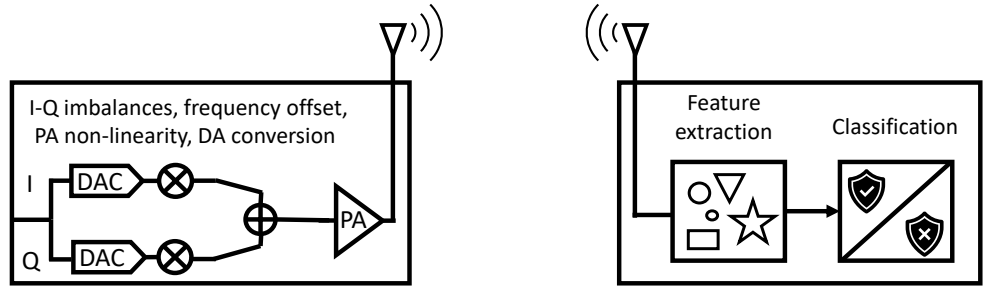
\begin{figure*}[!t]
\centering
\begin{tikzpicture}
\node[] at (0,0) {\includegraphics[clip, trim=2cm 2cm 6cm 9cm,width=0.8\textwidth]{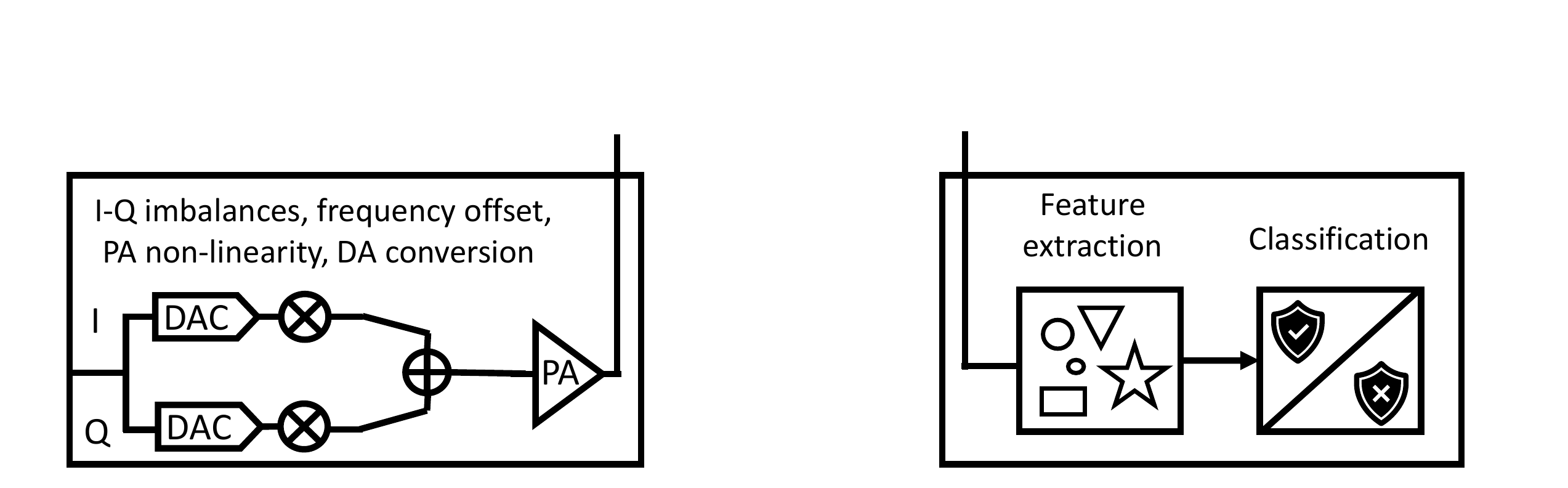}};
\draw[thick,radiation,decoration={angle=45}](1.56,1.3) -- +(180:0.55);
\draw[thick,radiation,decoration={angle=45}](-1.42,1.3) -- +(0:0.55);
\draw[ line width=0.5mm] (1.68,1) node[anchor=north]{}
  -- (1.53,1.4) node[anchor=north]{}
  -- (1.83,1.4) node[anchor=south]{}
   -- cycle;
\draw[ line width=0.5mm] (-1.54,1) node[anchor=north]{}
  -- (-1.39,1.4) node[anchor=north]{}
  -- (-1.69,1.4) node[anchor=south]{}
   -- cycle;
\end{tikzpicture}
\caption{RF-based authentication.}
\label{fig:RF}
\end{figure*}

RF-fingerprinting is the process of measuring the unique, stable and long-term imperfections of analog front-ends in wireless transceivers and wireless communication links~\cite{PLA_challenges}. Unlike PUFs, however, there is no guarantee of unclonability. 
Some of the  typically considered imperfections include in-phase quadrature-phase (IQ) imbalances, oscillator drifts, digital-to-analog conversion, power-amplifier non-linear characteristics, carrier frequency offset, etc.. The general idea of RF-based authentication is illustrated in Fig. \ref{fig:RF}, it is also explained briefly below. 

An RF-fingerprint based authentication protocol consists of two phases. First, an offline processing is carried out, where an authenticator captures a set of signals, extracts representative features and creates a classification function that maps features to a particular class, e.g., legitimate and not (optimally, the estimated features would perfectly describe all RF-imperfections of the transmitter). Next, during the online authentication phase, features of the received signals are measured and subsequently passed through the classifier (typically implemented as a hypothesis test). 

Location-based authentication relies on relating more specifically a node to a particular location. In detail, an authenticator should first obtain reliable information concerning the position of other nodes (e.g., a map that contains coordinates of other users). Next, the authentication process is based on online localization of users and comparing their estimated location to earlier stored coordinates. Authentication is successful if the estimated position passes a hypothesis test.

As an example the angle of arrival (AoA), time of arrival (ToA) and RSS-based methods~\cite{wifi_localization_survey} have all been used for localization-based authentication. In AoA based systems, receivers are expected to measure the angle of arriving signals, and, hence, must be equipped with an array of antennae. Depending on the number of available reference points, the angle information can be used towards direction or position finding. In ToA systems, to allow receivers measuring the distance to the transmitter time stamps are appended to the transmitted packets. To obtain accurate measurements, devices must have synchronized clocks. Finally, in RSS-based techniques, location can be determined by RSS measurements (i.e., channel fingerprints). 

A major advantage of both RF and location based authentication is that they enhance trust. Naturally, there are still challenges that must be addressed. Some of which are listed below. 
A major challenge comes from the increasing complexity in wireless systems, i.e., the chain of transmitter, channel and receiver. The two main approaches currently used to describe and optimize the system parameters are: i) model-based approach (e.g., using communication theory) where end-to-end communication is modelled as a set of blocks and each block can be parameterized and optimized independently, and, ii) model-free based approach (e.g., using machine learning techniques) where the whole system can be modelled and optimized as a single block~\cite{end-to-end-authentication}. The former approach is typically static, and, hence, performs well in stable environments. However, it could hardly capture the changes in dynamic environments. The latter approach have shown more success in complex environments but it usually requires great amount of training data and more computational power, hence, it is not well-suited for lightweight devices. 

The observations above indicate that a trade-off must be identified. First, the complexity of the environment (including number of devices, mobility, etc.) must be taken into account, i.e., devices should be able to adaptively switch between static and dynamic approaches. Recent practice leverages an initial approximate model and only uses training data to fine tune the representation. 

Secondly, the complexity of the approach should depend on the capabilities of the device.  
Overall, RF-based authentication would typically require high sensitivity at the receiver (e.g., spectrum analyzer) to identify the unique imperfections of the transmitter. Hence, RF-based authentication might be more suitable for unilateral authentication - access points to identify users. On the other hand, location-based authentication could be well suited for low-end devices, hence, when available could easily provide mutual authentication. 
Similarly, depending on the channel conditions (e.g., SNR, LoS or NLoS) and system parameters (number of subcarriers, antenae, etc.) RF-based authentication might be preferable over location-based, and vice versa. 

Another important issue concerns the accuracy of the collected location and RF information. Some of the factors that affect  accuracy are: choice of classification and loss functions, channel quality, choice of metrics and mobility of users. As discussed earlier both of the authentication approaches rely on pre-filled database (e.g., a channel model, a trained neural network or a downloaded map), however, during the authentication phase devices would observe noisy and time-varying features.

A promising research direction is online learning of the channel and feature selection aided by dimensionality reduction, to enable real-time analysis of multi-dimensional data~\cite{PLA_7}. 
Furthermore, a combination of features could also be considered for particular scenarios. 

Another topic that is gaining more attention due to the expected joint communication and sensing capabilities of 6G communication devices is waveform design. Finding a suitable waveforms that perform well for both communication and sensing would automatically improve authentication accuracy (e.g., of location-based methods) without causing additional overhead. 
Overall, continuous monitoring is required to adaptively change the authentication rules based on the variance of the collected features.

Next, one of the utmost important issues to be solved before adopting RF/location-based authentication approached is related to the threat model. While, all cryptographic based authentication protocols have a unified threat model, i.e., the well-known Dolev-Yao model\footnote{A Dolev-Yao type of adversary i) has control over the legitimate channel and can send any type of queries using knowledge that has been gained through observation of previous protocol executions; ii) all functions and operations used for authentication between the legitimate users are assumed public; iii) the adversary can perform denial of service (DoS) attacks to block parts of the authentication procedure and de-synchronize the legitimate connection.}, PHY-based techniques are typically based on different assumptions for the adversary. 

For example, location-based authentication might falsely identify an attacker as a legitimate user if the former is in the close vicinity to the latter. 
Therefore, as noted in the previous subsection a multi-factor authentication must be considered. Instead of relying on a single features, devices should capture, combine and classify based on multiple, i.e., this will make an attack harder as the attacker would need to impersonate all features. For RF fingerprinting methods, it is also important to mention that the underlying security features must be carefully chosen. 

Another issue is that studies would typically focus on a single attack, e.g., spoofing (multiple devices same ID), Sybil attacks (1 device multiple IDs), jamming or injection attacks, but there are only a few that consider all. This is a problem that clearly has to be addressed. 
It is important that a unified model that captures all threats present in wireless authentication is proposed. As noted earlier authentication should not be a static process and therefore, the model should involve variable parameters, e.g., depending on the frequency carrier and application different channel and location correlation should be assumed on possible eavesdroppers~\cite{PLA_7}.

Finally, we discuss possible use cases where RF/location-based authentication could be beneficial for the system's security. Access points and base stations are static. As a result, location could be easily introduced as an additional authentication factor to counteract on false base station attacks by using inverse localization (user locating the BS)~\cite{Mitev_Access_2022}. 

Furthermore, it could also allows APs and BSs to track devices, hence, an AP could easily predict the time when a device will leave a cell and enter a neighbouring one. Then, information could be transferred between APs to allow speeding up authentication for the device handover~\cite{PLA_challenges}. 

Another possible approach for identifying adversarial users could be through the uniqueness of antenna arrays in massive MIMO communication, e.g., the beam patterns from different devices will differ even if they are co-located. This can be used by authenticators to identify users in close proximity. A specific application is the sector level sweep (SLS) mechanism introduced in standards as IEEE 802.11ad for 60GHz mmWave WLAN.
SLS is the technique that identifies the beam pattern with optimal channel gains. In fact, it has has already been shown that the sweeping patterns could be used as unique and reliable source to counter spoofing attacks~\cite{SLS_Spoof}.

One of the main advantages of RF/location based authentication is that, both can provide per packet authentication. In this sense, it is clear that both techniques could, apart from authentication, contribute to methods such as anomaly detection and trust building.

The discussion above gives some initial ideas on how RF and location information could contribute to the system's security. However, there are still open issues that need to be addressed before their full integration into the standards. Depending on the application different problems may arise. For example, if considering a human-device the authentication would typically be end-to-end; if considering fully autonomous system authentication would be device-to-device~\cite{PLA_challenges}. An important research topic, for both cases, is the development of cross-layer security protocols. In particular, how should upper layers access, process and use PHY information. Answering this question could pave the way for a new lightweight and cross-layer security solutions.


\section{Future Perspectives: Context Aware 6G Security Incrorporating the PHY}

We are entering a new era of massive connectivity of autonomous cyberphysical agents, equipped with enhanced sensing, processing, and learning capabilities. In the respective networks, the devices involved will be highly heterogeneous (from RFIDs to autonomous vehicles), will be manufactured by different vendors without homogeneous processes and will operate under unprecedentedly diverse constraints (power, latency, computational and memory resources), all of which are challenging for security. Two further challenges that need to be taken into account include the introduction of AI and ML (6G will be the first native AI generation) and resistance against  quantum computers.

Confidentiality, integrity, accountability, access control and privacy must be at the core of today’s designs of future generations of intelligent networks. In the past, static security solutions were introduced as add-ons to earlier network design choices. A break from this paradigm is needed for the design of future security controls, primarily because:
\begin{itemize}
\item Static security solutions cannot scale efficiently while meeting latency, computation, power constraints, heterogeneity and lack of homogeneous production procedures;
\item AI/ML introduces novel vulnerabilities, the full extent of which is yet unknown;
\item It seems feasible for quantum computing to become commercially available.
\end{itemize}

\begin{figure*}[!t]
    \centering
    \includegraphics[clip, trim=0cm 2cm 0cm 0cm,width=0.8\textwidth]{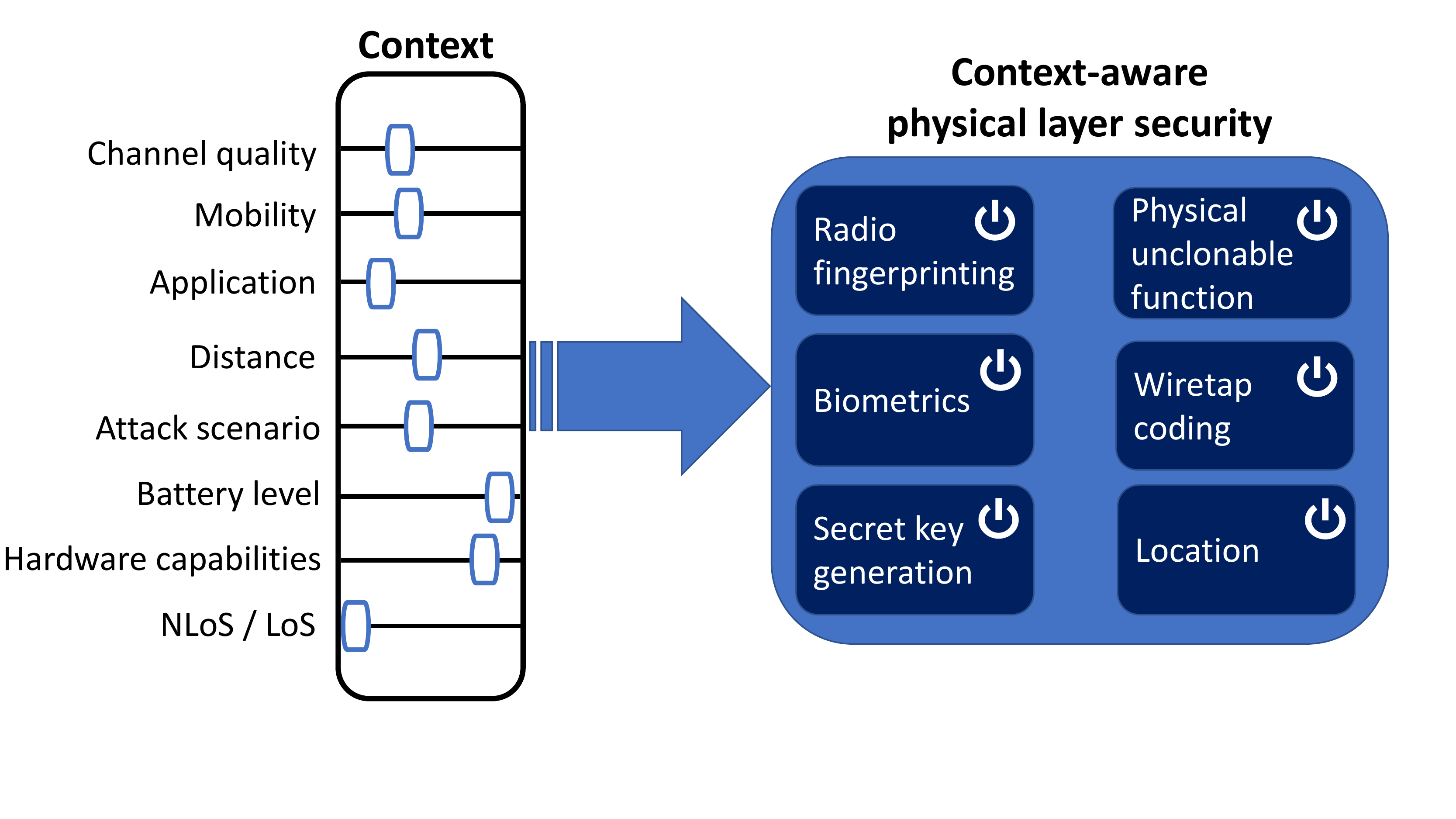}
    \caption{Context-aware PLS.}
    \label{fig:context_PLS}
\end{figure*}

The challenges ahead require a fundamentally different solution. To this end, looking beyond the current research horizon, a radically new, context-aware, AI-empowered roadmap for the design of intelligent and adaptive security controls is needed. This concept offers solutions to pressing needs, such as securing low-end IoT systems in the QoSec framework, or latency-constrained verticals, such as automotive and industrial IoT.
In addition, privacy should be addressed by design; both by means of new engineering solutions (e.g., for sensing, localization, private computations, private information exchange) and of course through a new legal and regulatory framework. 

In this framework, exploiting the characteristics of physical phenomena to provide security becomes pertinent. PLS can both complement conventional upper-layer security schemes and strengthen the overall trust and resilience of 6G. Different security solutions are attainable by exploiting  novel opportunities such as in sub-GHz to THz frequency bands, intelligent reflective surfaces, joint communications and sensing, localization and RF fingerprinting. Furthermore, the interplay between PLS and advances in artificial intelligence and machine learning, the role of semantics and context-awareness in the deployment of PLS based solutions needs to be considered. 

With this in mind, Fig. \ref{fig:context_PLS} showcases our vision towards a context-aware PLS. As discussed throughout this paper we do not believe that there exists a single PLS scheme that can be used in all possible scenarios. Instead, we think that a context-driven approach, which takes different PHY aspects into account, should be utilized. Depending on the available contextual information, PLS schemes could be used as lightweight security solutions, towards insuring trust.

It is clear that PLS is a set of useful tools which can greatly contribute towards the security of future networks. This article has presented our vision and some concrete examples on what PLS can do for the future generation of wireless networks. However, while there is a vast theory behind all PLS schemes, a generalized practical perspective is still missing. Along with all pros behind PLS we have also highlighted some of the major gaps in the area and hope that will stimulate further research.

\section*{Acknowledgement}

This work is financed by the Saxon State government out of the State budget approved by the Saxon State Parliament.

\bibliographystyle{IEEEtran}
\bibliography{bib}

\vspace{-10pt}

\begin{IEEEbiographynophoto}{Miroslav Mitev} received his B.Sc. (2014) and M.Sc. (2016) in communication and computer engineering from the Technical University of Varna, Bulgaria and Aalborg University, Denmark, respectively. In 2020, Miroslav obtained the Ph.D from the University of Essex, UK with his thesis in the area of physical layer security. Next (2020), he joined the École Nationale Supérieure de l'Électronique et de ses Applications (ENSEA) in France, where he was working in collaboration with Nokia Bell Labs on interference management. Since 2021, he is with the Wireless Connectivity group at the Barkhausen Institute, Germany. His research interests include wireless communications, physical layer security and link adaptation for industrial IoT.


\vspace{-10pt}
\end{IEEEbiographynophoto}

\begin{IEEEbiographynophoto}{Thuy M. Pham}
 (S’18) received the Ph.D. degree in wireless communications from Maynooth University, Ireland in 2020. In 2018, he worked for Airrays GmbH, Germany, as a part-time Research Engineer on an LTE Project. He is with Barkhausen Institut, Germany since 2020. His research interests include ad hoc wireless routing protocols, wireless communications, and physical layer security.
\end{IEEEbiographynophoto}
\vspace{-10pt}

\begin{IEEEbiographynophoto}{Arsenia (Ersi) Chorti}
is a Professor at the École Nationale Supérieure de l'Électronique et de ses Applications (ENSEA) in France, Joint Head of the Information, Communications and Imaging (ICI) Group of the ETIS Lab UMR 8051, Leader of the Wireless Connectivity Group at the Barkhausen Institut  in Germany and a Visiting Scholar at Princeton and Essex Universities. Her research spans the areas of wireless communications and wireless systems security for 5G / 6G, with a particular focus on physical layer security. She is a Senior IEEE Member, member of the IEEE INGR on Security and of the IEEE P1951.1 standardization workgroup (Smart Cities) and Chair of the IEEE Focus Group on Physical Layer Security. She is Associate Editor in Chief for the Nest reading of the Communications Society and Associate Editor of the IEEE Open Journal of the Signal Processing Society.
\end{IEEEbiographynophoto}
\vspace{-10pt}

\begin{IEEEbiographynophoto}{André Noll Barreto}
(Senior Member, IEEE) received the M.Sc. degree from the Catholic University (PUC-Rio), Rio de Janeiro, Brazil, in 1996, and the Ph.D. degree from the Technische Universität Dresden, Germany, in 2001, both in Electrical Engineering. He held several positions with academia and industry in Switzerland (IBM Research) and Brazil (Claro, Nokia Technology Institute/INDT, Universidade de Brasília, and Ektrum). From 2018 to 2022 he worked in the Barkhausen Institut, researching wireless communications for the reliable, resilient, and secure Internet of Things. He was the Chair of the Centro-Norte Brasil Section of the IEEE in 2013 and 2014, and the General Co-Chair of the Brazilian Telecommunications Symposium in 2012. Since September 2022 he is with Nokia Bell Labs in Paris, working on the evolution towards 6G.
\end{IEEEbiographynophoto}

\vspace{-10pt}
\begin{IEEEbiographynophoto}{Gerhard Fettweis}
 (Fellow, IEEE) is Vodafone Chair
Professor at TU Dresden since 1994, and heads the Barkhausen Institute since 2018, respectively. He earned his Ph.D. from RWTH Aachen in 1990. After one year at IBM Research in San Jose, CA, he moved to TCSI Inc., Berkeley, CA. He coordinates the 5G Lab Germany, and 2 German Science Foundation (DFG) centers at TU Dresden, namely cfaed and HAEC. His research focusses on wireless transmission and chip design for wireless/IoT platforms, with 20 companies from Asia/Europe/US sponsoring his research. Gerhard is IEEE Fellow, member of the German Academy of Sciences (Leopoldina), the German Academy of Engineering (acatech), and received multiple IEEE recognitions as well has the VDE ring of honor. In Dresden his team has spun-out sixteen start-ups, and setup funded projects in volume of close to EUR 1/2 billion. He co-chairs the IEEE 5G Initiative, and has helped organizing IEEE conferences, most notably as TPC Chair of ICC 2009 and of TTM 2012, and as General Chair of VTC Spring 2013 and DATE 2014.

\end{IEEEbiographynophoto}

\end{document}